\documentclass[12pt]{article}

\usepackage{graphicx}
\usepackage{epstopdf}
\usepackage[body={17.5cm, 21cm},right=2cm]{geometry}
\usepackage{amssymb}
\usepackage{amsmath}
\usepackage{psfrag}
\usepackage{epsfig}

\newcommand\nn{\nonumber}
\newcommand\ee{\end{equation}}
\newcommand\be{\begin{equation}}
\newcommand\eea{\end{eqnarray}}
\newcommand\bea{\begin{eqnarray}}

\def\ra{\rangle}
\def\beq{\begin{equation}}
\def\eeq{\end{equation}}
\def\d{\partial}

\begin{document}
\setcounter{page}{0}
\thispagestyle{empty}

\begin{titlepage}

\begin{center}

~

\vspace{1.cm}

{\LARGE \sc{
Microcausality in Curved Space-Time
}}\\[1cm]
{\large Sergei Dubovsky$^{\rm a,b}$, Alberto Nicolis$^{\rm a,c}$, \\[.2cm] Enrico Trincherini$^{\rm a}$, and Giovanni Villadoro$^{\rm a}$}
\\[0.6cm]

{\small \textit{$^{\rm a}$ Jefferson Physical Laboratory, \\ Harvard University, Cambridge, MA 02138, USA}}

\vspace{.2cm}
{\small \textit{$^{\rm b}$ Institute for Nuclear Research of the Russian Academy of Sciences, \\
        60th October Anniversary Prospect, 7a, 117312 Moscow, Russia}}

\vspace{.2cm}
{\small \textit{$^{\rm c}$ 
Department of Physics, Columbia University, \\
New York, NY 10027, USA }}

\end{center}

\vspace{.8cm}
\begin{abstract}
It is well known that in Lorentz invariant quantum field theories in flat space the commutator of space-like separated local operators vanishes (microcausality).
We provide two different arguments showing that this is a consequence of the causal structure of the classical theory, rather than of  Lorentz invariance.
In particular, microcausality holds in arbitrary curved space-times, where Lorentz invariance is explicitly broken by the background metric.
As illustrated by an explicit calculation on a cylinder this property is rather
non trivial at the level of Feynman diagrams.
\end{abstract}

\end{titlepage}

%%%%%%%%%%%%%%%%%%%%%%%%%%%%%%%%%

\section{Prelude: Microcausality in Flat Space}\label{prelude}

One crucial property of any consistent relativistic quantum field theory in Minkowski space is `microcausality'. Microcausality is the statement that the commutator of local (Heisenberg-picture) operators evaluated at space-like separated points vanishes,
\be
\big[ {\cal O}_1 (x),  {\cal O}_2 (y) \big] = 0 \qquad \mbox{for } \;
(x - y)^2 < 0 \; ,
\ee
and it is an exact, non-perturbative statement about the quantum theory.
Its relation to causality is obvious: according to quantum mechanics, if two operators commute measuring one cannot have any influence on measuring the other. 

Microcausality holds as an operator equation. However it is also instructive to consider its  vacuum expectation value (vev), since the vev of the commutator of two local operators describes the causal response of the vacuum to external sources. Indeed, it is the generalization of the retarded Green's function to an interacting theory.
Consider for instance a real scalar $\phi$ with generic self interactions, coupled to a local external source $J(x)$, ${H} = {H}_\phi + \int \!  d^3x \, J \phi$.
We can ask what is the vev of $\phi$ at some point $y=(y^0,\vec y)$ if we start with the vacuum at $t=-\infty$ and we then turn on a localized source $J$ for a finite period of time. 
In the interaction picture, treating $J\phi$ as the interaction, this is
\be
\langle  \phi(y)  \rangle_J = \langle 0 | \bar T e^ {i {\int_{-\infty} ^{y^0}} \!  dt \, d^3x \, J \phi}  \: \phi(y) \:  T e^ {-i {\int_{-\infty} ^{y^0}} \! dt \,  d^3x \, J \phi}  | 0 \rangle \; ,
\ee
where by $T$ and $\bar T$ we denote time-ordered and anti-time-ordered products, respectively. The above formula can be easily derived through Schwinger's `in-in' formalism (see ref.~\cite{weinberg_cosmo} for a nice review).
Expanding in the external source $J$ we see that at leading order the vev of the $\phi\phi$ commutator plays the role of a Green's function for $\phi$,
\be
\langle \phi(y)  \rangle_J = \int_{-\infty} ^{y^0} \!  d^4x \, J (x) \cdot i\,   \langle 0 | \, \big[ \phi(x), \phi(y) \big] \, | 0  \rangle + \dots \; 
\ee
If the $\phi\phi$ commutator vanishes outside the light cone, then the source can only generate a non-zero field inside its future light cone.
In the following by `microcausality' we will refer both to the more general operator statement and to its vacuum expectation value.

Microcausality is crucial in making the $S$-matrix Lorentz invariant. The $S$-matrix is constructed in terms of $T$-ordered products of local field operators. The $T$-ordering of fields evaluated at space-like separated points $x$ and $y$ is in principle not Lorentz invariant, since by a Lorentz boost one can invert the time-ordering of $x$ and $y$. A sufficient condition for recovering Lorentz invariance is that field operators at space-like distance commute. 

For a Lorentz-invariant theory microcausality can be proven in a variety of ways (see, e.g., ref.~\cite{weinberg}).
One proof makes use of the K\"allen-Lehmann representation. The two point-function can be written as a sum over intermediate states,
\be \label{KL}
\langle 0 | \, \phi(x) \phi(y) \, | 0  \rangle = \int_0 ^\infty \! d\mu^2 \rho(\mu^2) \, \Delta_+(x-y; \mu^2) \; ,
\ee
where $\mu^2$ is the intermediate states' invariant mass, $\rho(\mu^2)$ is the spectral density function, and $\Delta_+(x-y; \mu^2)$ is the {\em free} scalar two-point function for a particle of mass $\mu$:
\be
\Delta_+(x-y; \mu^2) = \frac{1}{(2\pi)^3} \int \! d^4 p \, \theta(p^0) \, e^{i p \cdot (x-y)}  \delta(p^2+\mu^2) \; .
\ee
Then the causal properties of the full quantum two-point function are the same as in a free theory. In particular
\be
\langle 0 | \, \big[ \phi(x), \phi(y) \big] \, | 0  \rangle =  \int_0 ^\infty \! d\mu^2 \rho(\mu^2) \,  \langle 0 | \, \big[ \phi(x), \phi(y) \big] \, | 0  \rangle  _{{\rm free, \, mass} = \mu} \; ,
\ee
which vanishes for space-like separated $x$ and $y$.

Also microcausality follows directly from canonical quantization of the theory. The system is quantized by imposing canonical commutation relations at $t=0$ and then evolving forward and backward in time through the unitary time-evolution operator defined by the Hamiltonian. Canonical commutation relations are preserved by the time-evolution. If Lorentz-boosts are a symmetry of the theory, they are represented in the Hilbert space of the theory through unitary operators. These too preserve the canonical commutation relations. This means that field operators at space-like distance will commute, since by a Lorentz-boost space-like separated points can be brought to lie on an equal time surface, where the corresponding field operators commute.

A less familiar argument which will be useful in the following exploits the Euclidean continuation of the theory. The vacuum expectation value of the $\phi \phi$ commutator is given by the imaginary part of the Feynman propagator,
\be \label{feynman}
\langle 0 | \, \big[ \phi(x), \phi(y) \big] \, | 0  \rangle = i \, {\rm sign}(x^0-y^0) \cdot {\rm Im} \,  \langle 0 | \, T \big\{ \phi(x) \phi(y) \big\} \, | 0 \rangle \; .
\ee
(More generally, $[\phi,\phi]$ as an operator is given by the anti-Hermitian part of $T\{\phi\phi\}$.)
In the Euclidean theory the Feynman propagator between two different points is real, being a path-integral of manifestly real quantities, 
\be
\langle 0 | \, T \big\{ \phi(0) \phi(x_E) \big\} \, | 0  \rangle = \int \!  {\cal D} \phi  \, e^{-S_E[\phi]} \, 
\phi(0) \phi(x_E) \; .
\ee
Of course we are assuming that the above path-integral can be made sense of by the removal of all UV divergences through the usual renormalization procedure.
In this case the only residual divergence we can have is in the limit $x_E \to 0$.
Now, the Euclidean theory is obtained from the Minkowskian one via the identification $t = i t_E$, with real $t_E$. If the theory is Lorentz invariant the propagator can only be a function of the invariant combination $t_E^2 + \vec x^2= - t^2 + \vec x^2$. Therefore the invariant statement is not simply that the Feynman propagator is real for real $t_E$---that is for $t_E^2 > 0$. Rather, it is that the Feynman propagator is real for positive $- t^2 + \vec x^2$, i.e.~outside the light cone. In other words, when we talk about the Euclidean continuation of a Lorentz invariant theory we are automatically talking about the region {\em outside} the light cone in Minkowski space. The Euclidean origin corresponds
to the whole light cone. The analytic continuation back to Minkowski space can be seen as a continuation {\em into} the light-cone---leading to a non-zero imaginary part of the propagator, as a result of the $i \epsilon$ prescription for circumventing the singularity localized on the light-cone ($=$ Euclidean origin).
As long as we stay outside the light-cone the Feynman propagator is manifestly real, and thus eq.~(\ref{feynman}) vanishes.

The point we want to stress here is that all these simple arguments assume, and crucially rely on the {\em exact Lorentz invariance} of the quantum theory. This is explicit in all the derivations we sketched above, with the exception of that based on the K\"allen-Lehman representation.
There too however Lorentz invariance is necessary in order to arrive to eq.~(\ref{KL}) in the first place, that is, to write the full propagator  as a sum of free propagators.
Also, without making explicit use of Lorentz invariance it is extremely hard to keep track of microcausality in the perturbative expansion---for instance by checking whether loop corrections to the Feynman propagator in real space give a non-zero imaginary part outside the light cone (see, e.g., the example in sect.~\ref{example}).

What happens then when we consider a local relativistic QFT in a fixed curved space-time? Such a theory is not Lorentz invariant. The fact that the classical Lagrangian is {\em locally} Lorentz invariant tells us that Lorentz invariance is going to be an approximate symmetry of the theory at high momenta/short distances. In particular the UV properties of the theory, like e.g.~renormalizability are not affected by the presence of a curved background (see e.g.~\cite{wald}). But certainly Lorentz invariance will not be an exact global symmetry of the full theory. Even for metrically flat but topologically non-trivial spaces (i.e., tori) there are calculable IR effects that break Lorentz invariance, the simplest example being the Casimir effect.
It is therefore not manifest whether microcausality is going to be a generic feature of QFTs in curved space-times, or whether instead it may be impaired by Lorentz-violating IR effects. Definitely all arguments mentioned above are not directly applicable to this case. Indeed the possibility that microcausality be generically violated in curved space-time has been the object of recent investigations \cite{hollowood1,hollowood2}, leading to the striking claim that QED itself at one-loop level features microcausality violations.

The purpose of the present paper is to show that microcausality {\em does} hold in curved space-time. We start in sect.~\ref{dispersion} by reviewing the connection between causality and analytic dispersion relations, such as the Kramers-Kronig relations between the real and imaginary parts of the refraction index in classical electrodynamics. We explain why the Kramers-Kronig relations are {\em not} applicable to quantum corrections in a QFT and why instead a more general set of dispersion relations have to be used in this case. These are of course very well known notions; the main point
we are stressing here is that to check the causal properties of a theory via analytic dispersion relations one needs to know the full {\em off-shell} structure of the propagator in
momentum space. One can easily get confused by applying indirect criteria based on notions of group velocity or Kramers-Kronig 
relations for the on-shell refraction index. Indeed, in the literature there is a rather extensive discussion on ``superluminal" effects
in QED, e.g. \cite{Drummond:1979pp,Scharnhorst:1990sr,Shore:2003zc,hollowood2}. 
To avoid these subtleties, we will focus directly on the commutator in coordinate space.

We then give two general arguments for microcausality in curved space-time, one based on the path-integral formulation of the theory, the other on canonical quantization (sect.~\ref{proof}). As we will see, the crucial result is that the causal structure of the full quantum theory is the same as that of the corresponding {\em classical} field theory. In particular, two local field operators will commute outside the light-cone whenever the classical theory cannot propagate information outside the light-cone. 
Notice that from this viewpoint the central role usually given to Lorentz invariance at the quantum level in proving microcausality in Minkowski space is somewhat misleading. Lorentz invariance may be crucial in ensuring the relativistic causal structure of the classical field theory, but from that point on microcausality is just a consequence of canonical quantization.

These arguments are implicitly known in the literature. For example the formulation of the black hole information paradox in terms of nice slices is closely connected with our path integral argument (see e.g.~\cite{Polchinski}). Still we felt it useful to give a dedicated discussion on the subject. 

The hypothesis that an interacting classical field theory does not propagate information outside the light cone is a highly non-trivial requirement, even in Minkowski space. A theorem due to Leray (see e.g.~\cite{wald2} and references therein) gives a general result for a system of PDE's that are linear in the fields' second derivatives and have the general form 
\be \label{leray}
G^{\mu\nu}(x; \phi_i, \nabla \phi_i) \, \nabla_\mu \nabla_\nu \phi_j = F_j(x; \phi_i, \nabla \phi_i) \; ,
\ee
where $G^{\mu\nu}$ is a globally hyperbolic effective metric, smooth function of the space-time point $x$ as well as of the fields and their first derivatives, $\nabla$ is any  derivative operator, and the $F$'s are smooth functions. The theorem states that in such a system the causal structure, i.e.~the causal dependence of solutions on initial conditions, is determined by the light-cones of the effective metric $G_{\mu\nu}$. Information propagates inside or along such light-cones. Notice that the effective metric in general depends on the field configuration, and so the causal structure will in general depend on the solution.
Now, for the classical relativistic Lagrangians that define usual renormalizable QFT's, the classical field equations are of the form (\ref{leray}) with $G_{\mu\nu}$ independent of the fields and their derivatives, and given simply by the (flat or curved) space-time metric $g_{\mu\nu}$. In this case the causal structure is given by the light-cones of $g_{\mu\nu}$, as expected. For non-renormalizable Lagrangians however, even if the field equations have the form (\ref{leray})  the effective metric $G_{\mu\nu}$ is generically different from $g_{\mu\nu}$, and superluminal propagation of signals may appear at the classical level about non-trivial field backgrounds, already in Minkowski space \cite{superluminal}. However a non-renormalizable Lagrangian does not lead to a well defined quantum theory, unless of course it admits a renormalizable UV completion, in which case such superluminal effects are absent.
We can therefore safely assume that the field theories we will be interested in have classically a well-posed initial value formulation with no superluminal propagation of signals.

\begin{figure}[t!]
\begin{center}
\includegraphics[width=12cm]{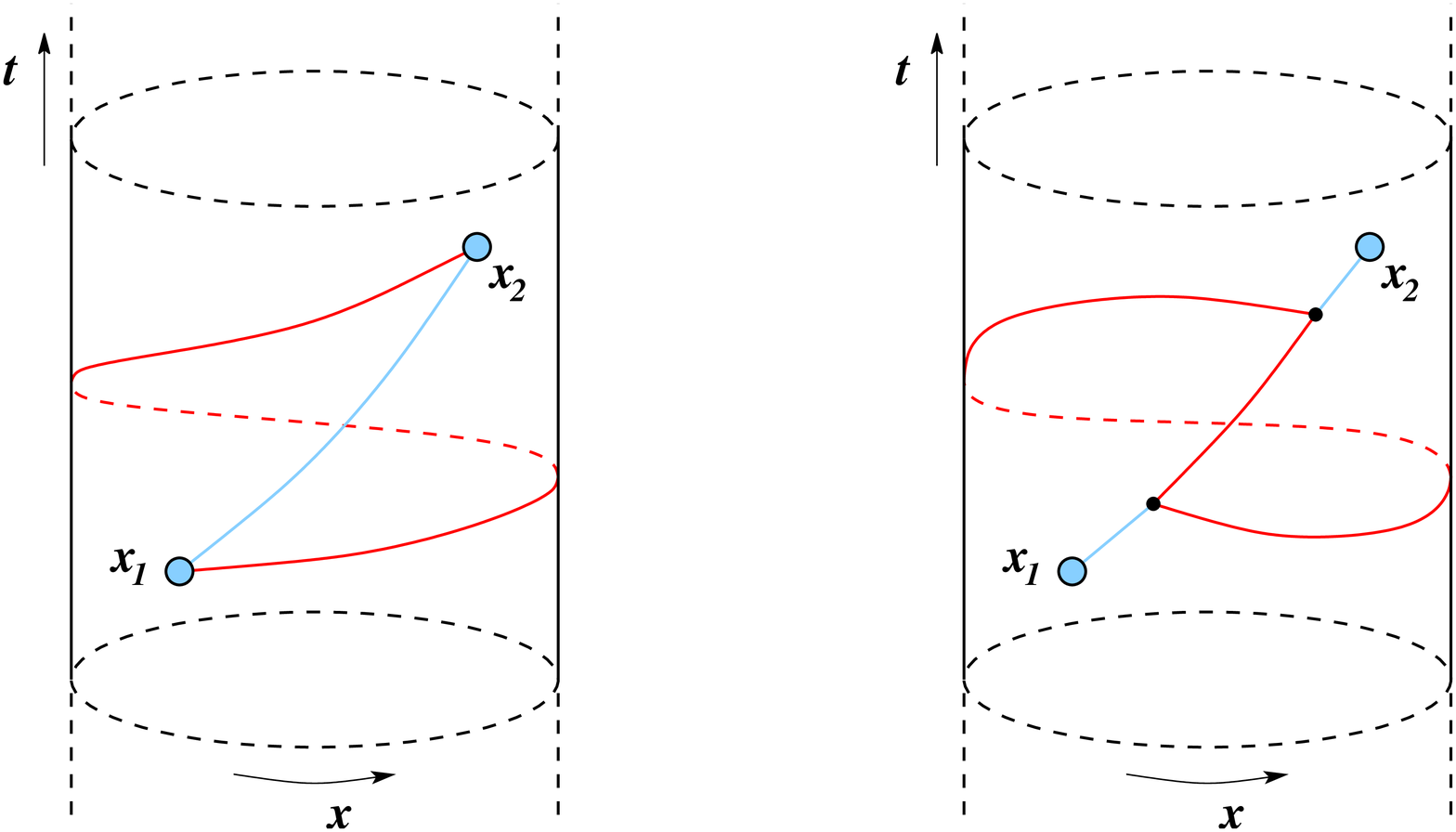}
\end{center}
\caption{\it \small \label{cylinder} {\em Left:} The tree-level propagator on the cylinder is the flat-space tree-level propagator summed over all images. The red line is the first topologically non-trivial contribution. {\rm Right:} At one-loop, apart from the sum over images of the one-loop propagators there are additional contributions coming from loops `wrapping' around the circle.}
\end{figure}

As an example of microcausality in Lorentz-violating space-times, we present an explicit one-loop computation in a non-trivial background (sect.~\ref{example}). To keep things as simple as possible, we consider a scalar field with $\phi^3$ self-interactions on a cylinder, i.e.~on a $D$-dimensional Minkowski space with one spatial dimension compactified on a circle.
Although such a space-time is metrically flat, the compactness of one spatial direction explicitly breaks Lorentz invariance, thus evading all traditional arguments reviewed above based on Lorentz invariance.
Indeed, at tree level the two-point function is the sum over images of the free two-point functions in Minkowski space
 (see fig.~\ref{cylinder}). 
%%%%%%%%%%%%%%%%%%%%%%%%%%%%
%%%%%%%%%%%%%%%%%%%%%%%%%%%%
%%%%%%%%%%%%%%%%%%%%%%%%%%%%
%%%%%%%%%%%%%%%%%%%%%%%%%%%%
Although not Lorentz invariant, it is at least causal in the sense that the retarded tree-level propagator vanishes outside the light cone---trivially, because the image contributions are not helping in going faster than light (going around the cylinder takes longer!). At one loop however, on top of the sum over images of the one-loop flat-space propagators, there are new contributions coming from loops wrapping around the circle. We will show that this effect cannot lead to any imaginary part of the Feynman propagator outside the light-cone. 

We present our conclusions in sect.~\ref{sum_rules}.

%%%%%%%%%%%%%%%%%%%%%%%%%%%%%%%%%
%%%%%%%%%%%%%%%%%%%%%%%%%%%%%%%%%
\section{Causality, Locality  and Analyticity\label{dispersion}}

It is well known that causality and subluminality (as we will discuss shortly,  in general these are 
different properties) manifest themselves as certain analytic properties of the Green's functions in momentum space.
To see how this connection arises, let us consider a field $\phi$ that satisfies an equation of the following general form
\be
\label{geneq}
F(i\d_t,i\d_x) \, \phi(t,x)=j(t,x) \; .
\ee
A generic solution to this field equation can be presented in the form
\[
\phi(t,x)=\int \! dt'dx'  \, G(t-t',x-x') \, j(t',x')\;,
\]
where $G(t,x)$ is a solution of the field equation (\ref{geneq}) with a delta-function source
at the origin $\delta(t,x)$.
A system described by (\ref{geneq}) is called {\em causal} if its Green's function can be chosen to satisfy the
boundary condition
\be
\label{causality}
G(t,x)=0 \qquad \mbox{for}\;\;t<0\;.
\ee
This property is just saying that the response of the system to a source cannot show up before
the source was turned on, and clearly should hold for all physical systems, both relativistic and non-relativistic. Note also that this discussion applies both to classical fields and to
the vev's of quantum fields. In the
latter case, as discussed in sect.~\ref{prelude}, $G(t,x)$ is the commutator $\theta(t) \, \langle[\phi(t,x),\phi(0,0)]\rangle$.

Causality imposes strong restrictions on the possible forms of the Green's function $G$
(and of the operator $F$). To see this  let us perform the Fourier transform of the Green's function with
respect to time, and define
\be
\label{timeint}
G(\omega,x)=\int \! dt\; e^{i\omega t} \, G(t,x) \; .
\ee
Causality (that is, eq.~(\ref{causality})) implies that the integration interval in (\ref{timeint}) is actually $t\in[0,\infty)$. If one furthermore assumes that the system is stable, {\it i.e.}~that the response to the source grows slower than an exponent with time, then $G(\omega,x)$ is analytic in the whole upper half complex $\omega$ plane, $\mbox{Im }\omega>0$. In principle, one can relax the last assumption and consider 
systems containing tachyonic degrees of freedom leading to exponential instabilities
with instability rate $\Gamma$. In order to be causal the Green's function of such a system has 
to be analytic at $\mbox{Im }\omega>\Gamma$. For simplicity, in what follows we assume that the system is stable.

In order to see the implications of {\em locality} let us perform also the Fourier transform with respect to the spatial variables $x$, and consider
\be
\label{fulltransform}
G(\omega,p)=\int \! dtdx\; e^{i\omega t+ipx} \, G(t,x)=F(\omega,p)^{-1} \; .
\ee
A very mild notion of locality implies that this Fourier transform exists at all real $p$, and combined with 
causality it results in the analyticity of $F(\omega,p)^{-1}$ as a function of $\omega$ at all real $p$ and
at $\mbox{Im }\omega>0$. Of course, usually by locality one understands much stronger restrictions on the properties of the system, which in turns restricts more the analytic 
properties of the function $F(\omega,p)^{-1}$. For instance, it is reasonable to assume that at each moment $t$ the response of the system to the delta-functional source is non-zero only in
a finite region of space. In this case $F^{-1}$ is an analytic function of both $\omega$ and $p$ at real $p$ and $\mbox{Im }\omega>0$. {\em Subluminality} is a stronger version of the last property which 
holds in all systems described by a relativistic quantum field theory. It says that $G(t,x)$ is causal and vanishes outside the light-cone, {\it i.e.}~when $t<|x|$. This translates into the analyticity of 
the function $F(\omega,p)^{-1}$ in the region $\mbox{Im }\omega>|\mbox{Im }p|$.

An important thing to keep in mind is that to check what the causal properties of a given system are in general one needs to know the whole {\it off-shell} function $F(\omega,p)$ describing the coupling to 
a local source as in eq.~(\ref{geneq}). It is not enough to know just the dispersion law for free waves, {\it i.e.}~the relation between $\omega$ and real $p$ such that $F$ is zero. To illustrate this important point
let us consider several specific examples. Let us start with a function $F$ of the following form
\be
\label{optical}
F(\omega,p)=n^2(\omega) \cdot \omega^2-p^2\;.
\ee
This kind of function appears, for instance, when one considers the propagation of light in a dispersive medium with $n(\omega)$ being the refraction index of the medium.
Indeed precisely in this context the relation between analyticity and causality was first exploited by Kramers and Kronig to derive their famous dispersion relations \cite{KK}.
It is instructive to see how the above $F(\omega,p)$ appears in a field theory language. Namely, a non-relativistic medium can be thought of as a set of decoupled localized degrees of freedom (atoms/molecules of the medium) which can interact with the photon.
For instance, approximating each degree of freedom as an harmonic oscillator of frequency $M$, one can write the following effective field theory action describing such a system
\be
\label{medium}
{\cal L}=\frac12(\d_\mu A)^2+\frac12(\dot X)^2-\frac12 M^2X^2-g A \dot{X}\;,
\ee
where $X$ is a field describing local degrees of freedom and $A$ stands for the photon field (for simplicity, we are suppressing all vector indices). By integrating out the $X$ field, one obtains the effective
action describing the propagation of the photon field with a kinetic function of the form (\ref{optical}), with 
\be
n^2(\omega) = 1+\frac{g^2}{M^2-\omega^2}\; .
\ee
This simple model actually provides an accurate description of the atomic contribution to the dielectric constant.

According to the above discussion, in order to describe a causal system the function  $F^{-1}$ should be analytic at all real $p$ and $\mbox{Im }{\omega}>0$. If $F$ has the form (\ref{optical}) this  condition implies
that the refraction index  $n(\omega)$ is also analytic. Indeed, if it has an essential singularity, then clearly $F^{-1}$ has an essential singularity as well. If $n(\omega)$ has a branching point $\omega_0$, then in order
for $\omega_0$ not to be a branching point of $F^{-1}$, this should be a square root branching, {\it i.e.} in the vicinity of $\omega_0$, one has either $n(\omega)\sim(\omega-\omega_0)^{1/2}$, or 
$n(\omega)\sim(\omega-\omega_0)^{-1/2}$. It is straightforward to check then, that in the first case, $F^{-1}$ has a pole in the vicinity of $\omega_0$ at small real $p$, while in the second
case it has a pole in the vicinity of $\omega_0$ at large real $p$. Analogously,  the existence of a pole singularity of $n(\omega)$ would lead to a pole of $F^{-1}$ at large real $p$.
So the requirement of causality imposes very strong constraints on a kinetic function of the form (\ref{optical}). In particular,  in this case causality almost implies subluminality; the only extra conditions  needed
are that at short scales   the dispersion relation takes the relativistic form, {\it i.e.}~$n\to 1$ when $\omega\to\infty$, and a certain positivity assumption.  Indeed, for the retarded propagator one has (for simplicity,
we are considering $(1+1)$-dimensional space-time)
\be
\label{retarded_optics}
G(t,x)=\int \! d\omega dp\; \frac{e^{-i(\omega t+px)}}{n^2\omega^2-p^2} \; .
\ee
Let us consider the case $x>0$; then by closing the contour of integration in the lower part of the $p$ plane and performing the integral over $p$ one obtains
\be
\label{omegaoptics}
G(t,x)=(2\pi i) \int \! \frac{d\omega}{2 \, n\, \omega} \; \sigma \, e^{-i(\omega t-\sigma n\omega x)}\;,
\ee
where 
\[
\sigma={\rm sign } {(\omega\,\mbox{Im }{n})}\;.
\]
If $\sigma$ is positive everywhere on the real axis, then it can be dropped from the integral (\ref{omegaoptics}). Then, at $x>t$ the asymptotic condition $n(\omega\to\infty)\to 1$ allows to close the contour
of integration in the upper half plane of $\omega$ and the analytic properties of $n$ discussed above imply that $G(t,x)$ is zero outside the light-cone. 

As an another example let us consider a kinetic function of the form
\be
\label{ghostial}
F(\omega,p)=\omega^2-f(p^2)\;.
\ee
Clearly, in many cases this function can give rise to the dispersion law for free waves as eq.~(\ref{optical}); however, as we will see now,
it has completely different causal properties. Indeed, this kinetic function corresponds to a second order differential equation in time, so that the
existence of a causal Green's function is automatic, and causality does not impose any constraints on the analytic structure of $f(p^2)$. As long as $f(p^2)$ is real and positive on the real axis, the kinetic function
(\ref{ghostial}) describes a stable causal system. However, as we are going to show now, unlike in the previous case the subluminality requirement is extremely powerful in this case---the only possible function $f$
that leads to a retarded propagator that vanishes outside the light-cone is the quadratic polynomial,
\[
f (p) = a \, p^2+b \, p+m^2 \; .
\]
To see this, it is convenient to use the subluminality condition discussed above, namely that $F(\omega,p)^{-1}$ is an analytic function of both its arguments in the region
$\mbox{Im }\omega>|\mbox{Im }p|$. This condition immediately implies that the function $f$ should be analytic in the whole complex plane. It is slightly non-obvious that $f$ cannot have a pole at some $p_0$, but it is straightforward
to check, that in the vicinity of the pole one could always find a point $p_*$ such that  $\mbox{Im }f(p_*)^{1/2}>|\mbox{Im }p_*|$, thus violating the subluminality condition. To complete the proof we will use only one extra assumption. Namely, we will make use of the fact that for reasonably local physical systems the Fourier transform
of the retarded propagator with respect to the spatial coordinate, which is equal to
\[
G(t,p)= \frac{\sin{ \big( f(p)^{1/2}t \big) }}{f(p)^{1/2}}
\] 
is an exponentially bounded function of the spatial momentum $p$. This implies that $f$ cannot grow faster than $p^2$ at infinity, and the only analytic function with this property is the quadratic polynomial.

We see that the two kinetic functions (\ref{optical}) and (\ref{ghostial}) describe systems with very
different causal properties, in spite of the fact that they can lead to identical on-shell
dispersion laws. As a very concrete example, let us consider
a kinetic function of the form
\be
\label{F1}
F_1=\omega^2-\frac{p^4}{p^2+1}\;.
\ee
According to the above discussion this kinetic function corresponds to a perfectly stable and causal system, which does not however possess a light-cone. The on-shell dispersion law 
following from this kinetic function is the same as that following from
\be
\label{F2}
F_2=\frac12(\omega^2+\omega\sqrt{4+\omega^2})-p^2\;.
\ee
This kinetic function has the form (\ref{optical}), and it does not describe a causal system due to the
presence of the branch cut in the upper half-plane of $\omega$. This somewhat counterintuitive result
is easy to understand. The point is that it does not matter which one of the functions
$F_{1,2}$ to put in eq.~(\ref{geneq}) as long as sources are absent, $j=0$: the free solutions are the same. However, the relation
between the two different sources that for the two different choices of the kinetic function create identical field configurations, is highly non-local. 
As a result the causal properties of the system, in its response to local sources are very different in the two cases. In other words, eq.~(\ref{geneq}), apart from providing a free wave dispersion law, carries a physically very important extra information---it describes how the system is coupled to local sources.
Without access to this information one cannot decide about the causal properties of the
system. 

Of course, the kinetic functions arising in relativistic quantum field theories have neither the form
(\ref{optical}), nor the form (\ref{ghostial}). According to the above discussion, in order to draw any conclusion about the causal structure of the theory from the direct study of  the analytic properties
of the kinetic functions, one needs to know the full {\em off-shell} result. For instance, a recent work
\cite{hollowood2} discussed the fate of (micro)causality in  QED in curved backgrounds.
At one loop the kinetic function in general has the form
\[
F=\omega^2-p^2-\alpha f(\omega^2,p^2)\;. 
\]
However, the authors of \cite{hollowood2} 
brought it into the form
(\ref{optical}) by means of the tree on-shell condition, and it was found that the resulting  $n(\omega)$ does not
enjoy the conventional analytical properties, {\it i.e.}~the Kramers-Kronig relations.
Due to the reasons explained above, this is not very surprising and by itself does not indicate a breakdown of (micro)causality. Actually, the necessity to know the full off-shell structure makes it somewhat unpractical to study the analytic properties of the propagator in momentum space to check microcausality in curved backgrounds in perturbative calculations. 
Instead, in what follows we will study the issue directly in position space.

%%%%%%%%%%%%%%%%%%%%%%%%%%%%%%%%%
%%%%%%%%%%%%%%%%%%%%%%%%%%%%%%%%%

\section{Microcausality in Curved Space\label{proof}}
In this section we will present two arguments that prove microcausality without relying on Lorentz invariance.
The arguments are very similar in spirit, but use somewhat different formalisms. Both of them show that the actual content
of microcausality is that the quantum theory inherits the causal structure present in the corresponding classical theory.

\subsection{Functional Integral Argument}
We will now show that the commutator of two operators ${\cal O}_{1,2}$ localized at any two points belonging to the same space-like hypersurface $\Sigma$ vanishes.
%%%%%%%%%%%%%%
%%%%%%%%%%%%%%
\begin{figure}[t]
\begin{center} \epsfig{file=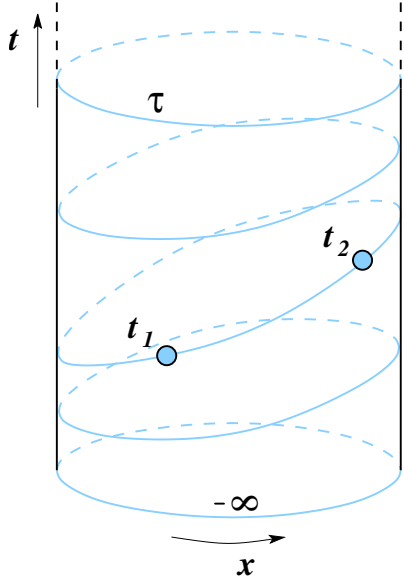,width=0.3\textwidth} \end{center} %,height=4cm
\vspace{-0.6cm}
\caption{{\it \small For any two space-like separated points at different times $(t_1, t_2)$ 
one can choose a set of coordinates such that they lie on the same time slice.}}
\label{fig:slices}
\end{figure} 
%%%%%%%%%%%%%%
%%%%%%%%%%%%%%
The essence of the argument is that one can quantize the theory by choosing the time coordinate
 in such a way that the surface $\Sigma$ corresponds to a constant time slice. With this choice of coordinates
 we are dealing with an equal-time commutator, $[{\cal O}_1(t),{\cal O}_2(t)]$, which is zero. Indeed, in canonical quantization
 \be
 \label{heisenberg}
 {\cal O}(t,x)=
 \bar{T}e^{{i\int^t_{-\infty}H}}\,
 {\cal O}(-\infty,x)\,Te^{{-i\int^t_{-\infty}H}}\,,
 \ee 
 so that
 \[
  [{\cal O}_1(t,x_1),{\cal O}_2(t,x_2)]= \bar{T}e^{{i\int^t_{-\infty}H}}\,[{\cal O}_1(-\infty,x_1),{\cal O}_2(-\infty,x_2)]\,
 Te^{{-i\int^t_{-\infty}H}}=0\;,
 \]
 where the last equality holds just by the definition of the canonical quantization. The only part of this argument that requires extra justification is the
 starting point---namely the statement that the result of the quantization does not depend on the choice of the time coordinate. To prove this desirable property it is convenient to use the functional integral representation of the matrix elements. Suppose that we quantized the system according to some time-slicing with time variable  $t$, and let us consider the matrix element 
\be
\label{matrix}
{\cal{M}}=\langle\psi_{1}|{\cal{O}}_1(t_1,x_1){\cal{O}}_2(t_2,x_2)|\psi_{2}\rangle\;,
\ee
where $|\psi_{1,2}\rangle$ are any two states.

By making use of (\ref{heisenberg}), which determines the time evolution of the Heisenberg operators, one can rewrite ${\cal M}$ in the
following form (for definiteness, we assume here that $t_1\geq t_2$, and the extension to the case $t_1\leq t_2$ is straightforward),
\be
\label{matrixelement}
{\cal M}=\langle\psi_{1}|\bar{T}e^{{i\int^\tau_{-\infty}H}}\,{T}e^{{-i\int_{t_1}^\tau H}}\,{\cal O}_1(-\infty,x_1)\,{T}e^{{-i\int_{t_2}^{t_1}H}}\,
{\cal O}_2(-\infty,x_2)\,{T}e^{{-i\int^{t_2}_{-\infty}H}}|\psi_{2}\rangle\;,
\ee
where $\tau$ is an arbitrary moment of time later than both $t_1$ and $t_2$, $\tau>t_{1,2}$. Now we can use the 
conventional representation of the (anti)time ordered products in the functional integral form and write (cf. e.g. \cite{weinberg_cosmo})
\begin{gather}
\label{fintegral}
{\cal M}=\int%_{\phi_L(T)=\phi_R(T)}
{\cal D}\phi_L\,
{\cal D}\phi_R \,
e^{-i\int_{-\infty}^\tau dt (L(\phi_L)-L(\phi_R) )} \, {\cal O}_1(\phi_R(t_1,x_1))\, {\cal O}_2(\phi_R(t_2,x_2))\,
\delta(\phi_L(\tau)-\phi_R(\tau))
%\Psi^*_1(\phi_L(-\infty))\Psi_2(\phi_R(-\infty))
\end{gather}
Here the integral over (left) right fields $\phi_{(L)R}$ represents the (anti)time ordered exponents and  the boundary conditions at $t=-\infty$ are determined by the states $|\psi_{1,2}\rangle$
in the matrix element (\ref{matrixelement}). For $t_1\leq t_2$ one obtains a similar formula, the only
difference being that operators ${\cal O}_{1,2}$ will depend now on the ``left" fields $\phi_L(t_{1,2})$.
The functional integral formula (\ref{fintegral}) provides a representation of the matrix elements that is totally independent of the 
choice of time slices used to quantize the system. Consequently, to calculate
the commutator in the operator formalism one can canonically quantize
the system using whatever choice of the time variable one finds convenient, as long as it provides a consistent Hamiltonian description. In particular in 
field theories  with two-derivative kinetic terms in curved backgrounds for any two operators belonging to a single space-like hypersurface
one can choose a time variable using this hypersurface as one of the slices, as we did above to show that the commutator vanishes.
Notice that this procedure fails if the two points are time-like separated---there can be no consistent 
Hamiltonian evolution connecting a space-like hypersurface with a time-like one. Indeed, in this case
somewhere in between the two slices  the $g^{00}$ component of the metric vanishes, so that the equations
of motion reduce to non-dynamical constraints and the Hamiltonian system degenerates
(see \cite{superluminal} for a related discussion).

%%%%%%%%%%%%%%%%%%%%%%%%%%%%%%%

\subsection{Canonical Quantization Argument}

Now we give another argument for microcausality in curved space-time entirely based on canonical quantization. 
We will show that in curved space-time the causal structure of a quantum field theory is the same as that of the corresponding classical theory. More precisely, if the classical theory one starts with cannot propagate information outside the light-cone, then upon canonical quantization of the theory commutators of field variables will vanish outside the light cone. 

Let's start with a generic mechanical system. Suppose we have a  classical Hamiltonian system with canonical variables $(q_1, \dots, q_N; p_1, \dots, p_N)$, collectively denoted by $(q,p)$, and Hamiltonian $H(q,p)$. At equal times the Hamiltonian variables obey canonical Poisson brackets, $\{ q_i(t),q_j(t) \}=0$, etc.
But what about Poisson brackets of variables at different times, for instance $\{ q_1(0), q_2(t)\}$? These depend on the dynamics. Indeed by solving the Hamilton equations
\be
\dot q_i = \{q_i, H \} \; , \qquad \dot p_i = \{ p_i,  H \}
\ee
one can express all canonical variables at time $t$ as functions of the initial conditions $(q(0), p(0))$ given at some earlier time $t=0$. So for example
\be \label{q2}
q_2 (t) = {\cal Q}_2 \big( t; q(0), p(0) \big) \; .
\ee
Then the Poisson bracket we are interested in is simply
\be \label{poisson}
\big\{ q_1(0), q_2(t) \big\} = \frac{\partial {\cal Q}_2 \big( t; q(0), p(0) \big)}{\partial p_1(0)} \; .
\ee
In particular, if for any dynamical reason the solution for $q_2$ at time $t$ does not depend on the initial condition $p_1(0)$, then the Poisson bracket between $q_1(0)$ and $q_2(t)$ will vanish.

We now quantize the system by promoting all Hamiltonian variables to operators---which we will denote by `hatted' variables $\hat q$, $\hat p$---and imposing canonical commutation relations. In Heisenberg picture we can ask again what is the commutator of canonical variables at different times, e.g.~$\big [ \hat q_1(0), \hat q_2(t) \big]$. 

The Hamilton equations now read
\be
i \frac{d}{dt} { \hat q_i} = \big[ \hat q_i,  H(\hat q, \hat p) \big] \; ,
\qquad 
i \frac{d}{dt} \hat p_i = \big[ \hat p_i,  H (\hat q, \hat p) \big] \; .
\ee
By definition the quantum Hamiltonian $ H (\hat q, \hat p)$ has the same functional dependence on the operators $\hat q$, $\hat p$ as the classical Hamiltonian has on the classical canonical variables $q$, $p$. The  only difference resides in possible ordering ambiguities, which are absent in the classical theory whereas in the quantum one they can lead to physically different systems.
Also the algebra of canonical commutation relations in the quantum theory is the same as the algebra of ($i \, \times$) canonical Poisson brackets in the classical theory. As a result of these facts the quantum canonical variables $\hat q, \hat p$ satisfy operator Hamilton equations that are functionally the same as the classical ones, again apart from ordering issues. This property is often referred to as `the quantum variables satisfy the classical e.o.m.~as operator equations'.
As a consequence the operator solutions to the quantum Hamilton equations will be functionally the same as the classical solutions, in their $t$-dependence as well as in their dependence on the canonical variables at some earlier time $t=0$. So for example the operator $\hat q_2$ at time $t$ will be
\be
\hat q_2 (t) = {\cal Q}_2 \big( t; \hat q (0), \hat p (0) \big) \; ,
\ee
where ${\cal Q}_2$ is {\em exactly the same function appearing in the classical theory}, eq.~(\ref{q2}), with some properly chosen ordering of the $\hat q$'s and $\hat p$'s.
We now can use the {\em classical} solutions to compute commutators of variables at different times in the {\em quantum} theory. So for example
\be \label{commute}
\big [ \hat q_1(0), \hat q_2(t) \big] = \big[  \hat q_1(0) , {\cal Q}_2 \big( t; \hat q (0), \hat p (0) \big)   \big] \; , 
\ee
where ${\cal Q}_2$ is the classical solution for $q_2(t)$, with some specific ordering. Now, the point is that if for any reason in the classical theory the solution for $q_2$ at time $t$ is not sensitive to the value of $p_1$ at $t=0$, then the above commutator will vanish. This is because $\hat p_1(0)$ is the only canonical variable at $t=0$ with which $\hat q_1(0)$ does not commute. But if the {\em classical} solution for $q_2(t)$ does not involve $p_1(0)$, the {\em quantum} solution for $\hat q_2(t)$ will not involve the operator $\hat p_1(0)$. In this case there will be no $\hat p_1(0)$ in the right argument of the commutator (\ref{commute}), and the commutator itself will vanish. This is obviously the quantum analogue of eq.~(\ref{poisson}). 
Notice that in this respect ordering ambiguities are harmless: they cannot change whether the solution depends or not on some of the initial conditions.
 
It is now clear where we are headed.
All the discussion above applies unaltered to a field theory in curved space-time, which is nothing but an Hamiltonian system with infinitely many degrees of freedom---one at each point in space. For any given time-slicing of the space-time manifold, we call $\phi_{\vec x}$ the classical field variables and $\Pi_{\vec x}$ their conjugate momenta, with $\vec x$ being a set of spatial coordinates parameterizing the time slices. By solving the Hamilton equations one can in principle express the field at time $t$ and position $\vec x$ as a functional of the initial field and momentum configurations given on some earlier time slice, for instance at $t=0$,
\be \label{classical_field}
\phi_{\vec x}(t) = \Phi_{\vec x} \big(  t; \phi(0), \Pi(0) \big) \; ,
\ee
where by $\phi(0)$, $\Pi(0)$ we denote the field and momentum configurations on the whole $t=0$ slice. Then, as in eq.~(\ref{poisson}), the Poisson bracket of field variables at different times is given by
\be  \label{poisson_field}
\big\{ \phi_{\vec x}(0) , \phi_{\vec y}(t) \big\} = \frac{\delta \Phi_{\vec y} \big(  t; \phi(0), \Pi(0) \big)}{\delta \Pi_{\vec x} (0)} \; ,
\ee
where $\delta (\cdots) / \delta (\cdots) $ is a functional derivative.

\begin{figure}[t!]
\begin{center}
\includegraphics[width=9cm]{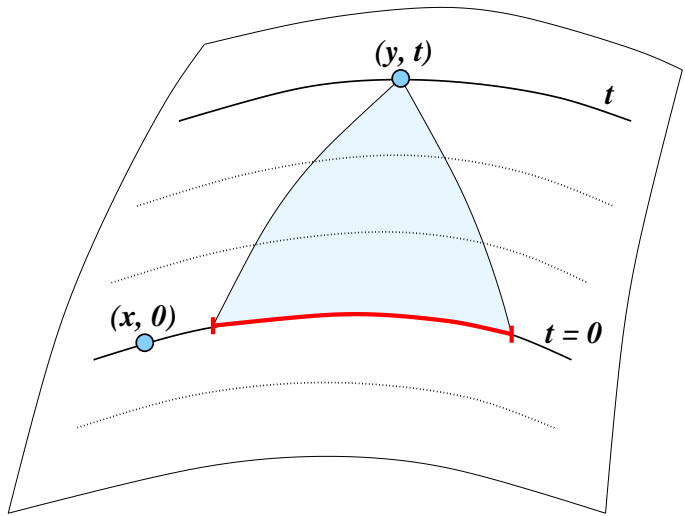}
\end{center}
\caption{\it\small \label{general_figure} In a classical theory that does not admit superluminal propagation of signals, the field at $(\vec y, t)$ can only depend on initial conditions given inside the point's past light cone {\em (shaded region)}. In particular, it cannot depend on the field and momentum values at $(\vec x,0)$.}
\end{figure}

The system is quantized by imposing canonical commutation relations. We are then interested in the commutator of field operators at different space-time points, e.g.~$\big [ \hat \phi_{\vec x} (0), \hat \phi_{\vec y} (t) \big]$.
Then in this case eq.~(\ref{commute}) reads
\be \label{commute_field}
\big [ \hat \phi_{\vec x} (0), \hat \phi_{\vec y} (t) \big] = \big[   \hat \phi_{\vec x} (0) , {\Phi}_{\vec y} \big( t; \hat \phi (0), \hat \Pi (0) \big)   \big] \; ,
\ee
where $\Phi_{\vec y}$, is the same functional appearing in the classical solution, eq.~(\ref{classical_field}).
Now,
\begin{itemize}
\item[{\em (i)}] if the points $(\vec x, 0)$ and $(\vec y, t)$ are space-like separated, and 
\item[{\em (ii)}] if the classical theory cannot propagate information outside the light cone, 
\end{itemize}
then the {\em classical} field at $(\vec y, t)$ cannot depend on the initial conditions given at the point $(\vec x, 0)$. It can only depend on initial conditions given inside the past light cone departing from $(\vec y, t)$, see fig.~\ref{general_figure}.
Hence, the functional $\Phi_{\vec y} \big(  t; \phi(0), \Pi(0) \big)$ does not depend on $\Pi_{\vec x} (0)$. This leads to the vanishing of the classical Poisson bracket eq.~(\ref{poisson_field}); but it also leads to the vanishing of the commutator eq.~(\ref{commute_field}) in the {\em quantum} theory, 
\be
\big [ \hat \phi_{\vec x} (0), \hat \phi_{\vec y} (t) \big]  = 0 \; ,
\ee
since $\hat \Pi_{\vec x} (0)$ is the only canonical variable at $t=0$ with which
$\hat \phi_{\vec x}(0)$ does not commute.
In this precise sense the causal structure of the quantum theory is the same as that of the classical one.

\vspace{24pt}
These two arguments can be straightforwardly extended to gauge theories as well.
The key is to work in a gauge with a well-defined Hamiltonian evolution.
For instance, in the first argument one can  start by defining the matrix element 
(\ref{matrix}) in the $A_0=0$ gauge, or in covariant form
\be
\label{nmugauge}
A_\mu n^\mu=0\;,
\ee
where $n^\mu$ is the time-like normal vector to the selected set of slices. In this gauge, one obtains an Hamiltonian system with canonical
variables $(A_i,F_{0i})$ and a Gauss's law constraint ${\cal C}$, which generates the residual
gauge transformations. Upon quantization this constraint is used to define the physical states
 $|\psi\ra$'s
 by imposing the condition
\be 
\label{phys_cond}
{\cal C}|\psi\ra=0\;.
\ee
Now one can construct the functional integral representation of the matrix elements. The only difference
with the case discussed above is that naively this functional integral depends on the choice
of time slices through the vector $n^\mu$. However,  integrals with different $n^\mu$'s differ only in the 
choice of the gauge fixing condition (\ref{nmugauge}) so that by making use of
the conventional Faddeev-Popov trick one can show that the matrix elements between physical
states do not depend on $n^\mu$. Consequently,  physical matrix elements do not depend on the
choice of the time slices used to quantize the system, and one can run the argument as before.

Similarly, one can use the same gauge to run the analogue of the second argument. In the
classical theory all solutions with the same initial data differ just by a gauge transformation and,
in particular, they are described by pure gauge configurations outside the light cone.
Correspondingly, following the same logic as above,  Poisson brackets (and commutators in the
quantum theory) are pure gauge outside the light cone. Then the condition (\ref{phys_cond})
implies that their matrix elements between physical states vanish.

Finally, it is worth mentioning that it is straightforward to include fermions too, for instance the first argument
applies to systems with fermions without any modifications.

To conclude this section, it is worth stressing that these two arguments can be applied in more general contexts because
they do not rely at all on the Lorentz invariance of the theory in flat background.
They will go through for any non-relativistic theory allowing an Hamiltonian description with a well 
defined causal structure at the classical level. As we already
stressed their real message is that the causal structure of the quantum theory 
is inherited from the classical theory without any modifications.
For example in non-Lorentz invariant theories with two derivative Lagrangians 
the commutator will vanish outside the lightcone of the field with the largest velocity.
On the other hand in theories with two time derivatives and higher order spatial derivative terms
one can see that subluminality is lost both from the analiticity discussion in sect.~\ref{dispersion}
and from the fact that with a non-trivial choice of time slices also higher time derivative terms 
are produced preventing the definition of an Hamiltonian evolution.

%%%%%%%%%%%%%%%%%%%%%%%%%%%%%%%%%

\section{Perturbative Example: $\lambda\phi^3$ on the Cylinder\label{example}}

The arguments we gave in the previous section, although general and very intuitive, hide
the non-triviality of the result in general non Lorentz invariant backgrounds.
In particular they do not rely on perturbation theory and Feynman graphs, 
the reason is simple---the cancellation of the commutator outside the light cone
is everything but manifest when expressed in terms of Feynman diagrams.
In order to show this point and to give a independent check of the arguments
we present here an explicit one-loop calculation of the commutator 
in a non Lorentz invariant background. 
Consider $\lambda \phi^3$ theory on a cylinder in $D$ space-time dimensions 
(namely on ${\mathbb R}^{D-1}\times S^1$). 
At tree level the 2-point function in coordinate space is just the sum over images of
the Lorentz invariant free-propagator whose imaginary part vanishes outside the light cone, namely
\bea
G^{\rm free}_{R}(x,y)&=&\sum_{n=-\infty}^\infty G^{\rm free}_\infty(x,y+R n) \,, \nn \\
G^{\rm free}_\infty(x,y)&=&\frac{\mu^{D-2}}{(2\pi)^{D/2}} \frac{K_{D/2-1}(\mu\, \sqrt{x^2+y^2})}{(\mu\, \sqrt{x^2+y^2})^{D/2-1}}\,,
\eea
where $\mu$ is the mass, $R$ the size of the circle, $x$ and $y$ the coordinates over $\mathbb R^{D-1}$ and $S^1$ respectively
and $K_{\nu}(s)$ is the modified Bessel function.
Since the commutator vev is proportional to the imaginary part of the Feynman propagator (see eq.~(\ref{feynman})),
the former trivially vanishes too. At higher loop orders, however, the 2-point function on the cylinder
is not just the sum over images of the 1-loop 2-point functions on the plane, since it receives also
contributions from loops wrapping the circle (see fig.~\ref{cylinder}). 
We present here the calculation of the commutator at one-loop order to show how non-trivial the cancellation is. 
We do not need to calculate explicitly the 2-point function, it is enough to show that its imaginary part vanishes. Should we start the calculation directly in Minkowski signature we would encounter many complex
contributions, and for the imaginary part to vanish one needs highly non-trivial cancelations between them.
So it is highly advantageous  first to perform the computation in Euclidean space (where the 2-point function is real by definition)
and then show that one can analytically continue the result to Minkowski space-time. 
The one-loop correction to the two-point function in position space is
\bea
\delta_{\rm 1-loop} \,G_R(x,y)=\lambda^2\, \int d^{D-1} p \sum_{p_y} \frac{ \Pi(p,p_y)}{(p^2+p_y^2+\mu^2)^2}\,e^{ipx+i p_y y}\,, \nn
\eea
where here and in the following we suppressed all numerical coefficients and $\Pi(p,p_y)$ is the 
one-loop integral, which after the introduction of the Feynman parameter reads
\bea
\int_0^1 d\xi \int d^{D-1} k \sum_{k_y} \frac{1}{[k^2+(k_y-\xi p_y)^2+\mu^2+(p^2+p_y^2)\xi (1-\xi)]^2}\,. \nn
\eea
Applying the Poisson summation formula
\bea
\sum_{n=-\infty}^{\infty} f(n)=\int_{-\infty}^{\infty} d\tau \sum_{m=-\infty}^{\infty} f(\tau) e^{i 2\pi m \tau}\,,\nn
\eea
to the $k_y$ sum we get
\bea
\Pi(p,p_y)=\int_0^1 d\xi \int d^D k \sum_{m} \frac{e^{i m (k_y+p_y\xi)R}}{[k^2+(k_y)^2+\mu^2+(p^2+p_y^2)\xi (1-\xi)]^2}\,.\nn
\eea
Applying again the Poisson formula to the $p_y$ sum we arrive at the following expression for the propagator
\bea
\delta_{\rm 1-loop} \,G_R(x,y)=\lambda^2\, \int_0^1 d \xi \int d^D p \int d^D k \sum_{n,m} \frac{e^{ipx}e^{ip_yy}}{(p^2+p_y^2+\mu^2)^2} 
	\frac{e^{im(k_y+p_y \xi) R +i n p_y R}}{[k^2+k_y^2+\mu^2+(p^2+p_y^2)\xi(1-\xi)]^2}\,.\nn
\eea
After evaluating the integral over $k$ by introducing the Schwinger parameter we get
\bea
\int_0^1 d \xi \int d^D p \sum_{n,m} \frac{e^{ipx}e^{ip_y (y+n R+\xi m R)}}{(p^2+p_y^2+\mu^2)^2} 
	 (|m| R)^{4-D} \frac{K_{2-D/2}\Bigl(|m| R \sqrt{\mu^2+(p^2+p_y^2)\xi(1-\xi)}\, \Bigr)}
	 {\left[|m| R \sqrt{\mu^2+(p^2+p_y^2)\xi(1-\xi)}\, \right]^{2-D/2}}\,. \nn
\eea
Because the integral in 
$p$ and $p_y$ is the Fourier transform of a function that depends
only on $p^2+p_y^2$, the propagator will be a sum of terms that are functions of the combination 
$x^2+\tilde y^2$, where $\tilde y\equiv y+nR+\xi m R$. 
Namely
\be \label{eq:GofF}
G_R(x,y)=\lambda^2\, \int_0^1 d \xi \sum_{n,m} F_m(x^2+\tilde y^2)\,.
\ee 

By definition the propagator in Euclidean space (\ref{eq:GofF}) is real.
At $x^0=0$ the Euclidean and Minkowskian 2-point functions coincide, so the 
Minkowski propagator is also real.  As long as the analytic continuation of the propagator 
to Minkowski space is free from divergences also the propagator in Minkowski 
space will be real and the commutator will vanish.

%%%%%%%%%%%%%%
%%%%%%%%%%%%%%
\begin{figure}[t]
\psfrag{Ap}{$A'$} \psfrag{Bp}{$B'$} \psfrag{A}{$A$} \psfrag{B}{$B$}\psfrag{t}{$t$} \psfrag{ytilde}{$\tilde y$} 
\begin{center} \epsfig{file=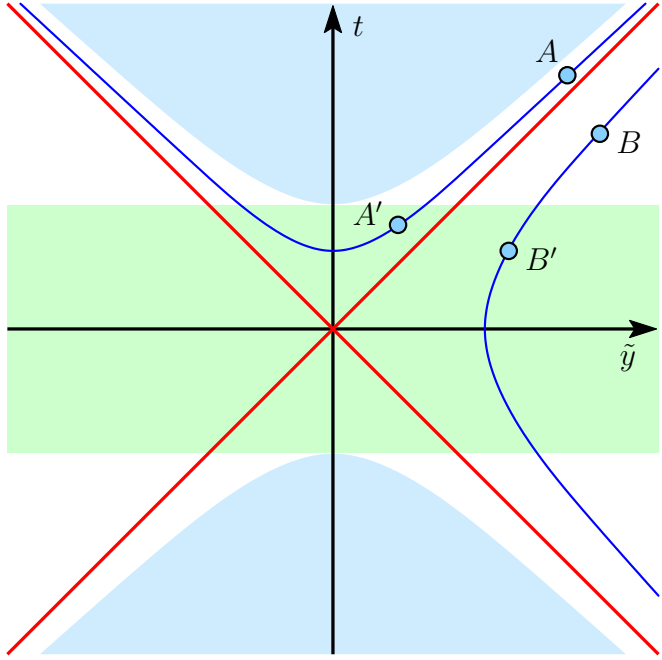,width=0.5\textwidth} \end{center} %,height=4cm
\vspace{-0.6cm}
\caption{{\it \small Plot of the $(t,\tilde y)$ plane. The shaded hyperbolic regions correspond to points inside the light cone in the $(t,y)$ plane
(eq.~(\ref{eq:tylessmR})). The shaded stripe is the region where the Wick-rotated $F_m$ integral converges (eq.~(\ref{eq:tlessmR})).
The blue hyperbolas are constant $t^2-\tilde y^2$ curves. Therefore for every point outside the light cone
(e.g.~$A$ or $B$) there exists a point (e.g.~$A'$ or $B'$, respectively) with the same value of $t^2-\tilde y^2$ (and thus of $F_m$)
such that the integral converges.}}
\label{fig:ineq}
\end{figure} 
%%%%%%%%%%%%%%
%%%%%%%%%%%%%%

%After the Wick rotation the exponent $\exp(i p x)$ in eq.~(\ref{eq:GofF}) goes to 
%$\exp(-p_0 t+i\vec p \vec x)$. The integration over $p_0$ could now diverge for large $|p_0|$ 
%and give raise to an imaginary part. Now we want to show that for points outside the light cone 
%the convergent Euclidean integral can be analytically continued safely.
Notice that outside the light cone the following inequality holds:
\be \label{eq:tylessmR}
-t^2+\tilde y^2=-t^2+(y+nR+\xi m R)^2>-m^2 R^2 \xi(1-\xi)\,.
\ee
Indeed, the quadratic term in $\xi$ is the same on both sides, and it is then trivial 
to check that the inequality holds outside the light-cone at the boundary values $\xi=0,1$. 
Points inside the light cone thus correspond 
to the hyperbolic shaded regions in fig.~\ref{fig:ineq}. 
For $-t^2+\tilde y^2>0$ no analytic continuation is needed---analogously to the Lorentz invariant case
$F^{\rm Mink}_m(t,\tilde y)\equiv F_m(-t^2+\tilde y^2)$ and its imaginary part vanishes. 
For $-m^2 R^2 \xi(1-\xi)<-t^2+\tilde y^2<0$ the proof needs one more step.
If we analytically continue $x^0\to i t$
the integrand in $F_m$ for large $p_0$ becomes
\bea
\frac{(m R)^{4-D}\ e^{-p_0 t+i\vec p \vec x+ip_y \tilde y -|m| R \sqrt{\mu^2+(p^2+p_y^2)\xi(1-\xi)}}}
{\Bigl(p^2+p_y^2+\mu^2\Bigr)^2\, \Bigl(|m|R \sqrt{\mu^2+(p^2+p_y^2)\xi(1-\xi)}\, \Bigr)^{(5-D)/2}}\,, \nn
\eea
in particular the leading exponential contribution reads
\bea
e^{-p_0 t -|m| R |p_0|\sqrt{\xi(1-\xi)}} \,. \nn
\eea
When the integral is convergent, i.e. for\footnote{We are only interested in 
the $m\neq0$ case, since for $m=0$ the 2-point function is the sum over images of the
Lorentz invariant 1-loop contributions, which respect microcausality.}
\be \label{eq:tlessmR}
t^2<m^2 R^2 \xi(1-\xi)\,,
\ee
$F^{\rm Mink}_m(t,\tilde y)\equiv F_m(-t^2+\tilde y^2)$ is real and Lorentz invariant with respect to
the vector $(t, \tilde y)$. We can thus analytically continue $F^{\rm Mink}_m$ everywhere outside the
light cone (eq.~(\ref{eq:tylessmR})) using this invariance. 

Therefore, for every point outside the light cone $F_m$ (and thus $G_R$) 
can be analytically continued without incurring into singularities, 
so that the Minkowski propagator is real and the commutator vev vanishes.
We see that microcausality is indeed a highly non-trival property at the level of the Feynman diagrams;
even in this relatively simple example one needs to follow the structure of the loop integrals at a detailed level
(such as the precise way in which the Feynman parameter $\xi$ enters into different expressions) to see the required cancellations.

%%%%%%%%%%%%%%%%%%%%%%%%%%%%%%%%%

\section{Discussion\label{sum_rules}}

While Lorentz invariance is not necessary for microcausality to hold, it is usually
exploited in the standard proofs because it makes the vanishing of commutators outside the light-cone manifest.  
We gave two general arguments that show why also in non-Lorentz invariant backgrounds
microcausality is expected to hold. The first makes use of path integrals and their manifest
independence on the coordinate choice. In this way two space-like separated
points can be chosen to lie on the same time-slice. This fact together with the existence of a unitary 
evolution in any given time-slicing ensures microcausality. The second argument instead
is basically the statement that quantization is the procedure that maps functions of
classical variables satisfying canonical Poisson bracket relations into the same  functions of operators 
satisfying canonical commutation relations. In this way points that are causally disconnected
at the classical level (vanishing Poisson brackets) are automatically causally disconnected
also at the quantum level (vanishing commutators).
Let us discuss some potential subtleties in the above two arguments.
First, in quantum field theory it may happen that, due to the  
presence of the infinite number of degrees of freedom, 
the commutators of operators are modified as compared to the classical Poisson brackets,
the most famous examples being the anomalous violation of the chiral  
symmetries and the central extension of the Virasoro algebra in string theory. In the  
functional integral argument this subtlety translates into the necessity to define a properly  
regularized measure that does not modify the causal structure of the  non-regularized theory. 
We do not expect this subtlety to be relevant in the case under consideration, as the  
quantum anomalies arise due to UV effects and exhibit themselves as  
localized terms in position space, while we are discussing commutators of space-like separated  
operators.  For instance,  dimensional regularization in curved space-time \cite{Luscher:1982wf} does not modify the causal  
structure of the theory, although care should be taken when  applying it in
Lorentzian signature (other popular covariant methods of regularization, such as point-splitting or adding higher  
derivative terms in general do modify the causal structure of the theory).

Another subtlety (present only in the functional integral proof) is  
that it is not totally obvious that the impossibility to connect  
two points by a time-like or null curve (which is the most natural
definition of space-like separation) automatically implies that there exists a smooth space-like Cauchy surface on which the two points both lie. 
It is easy to check that  
this is true for some simple space-times and would be interesting  
to understand under what conditions these two properties are equivalent.

We also reported an explicit one-loop calculation of the commutator in a non-trivial background,
which shows that even in the simple case where Lorentz invariance is broken just by boundary conditions,
checking microcausality in the perturbative expansion may be rather cumbersome. The fact that the vanishing 
of the commutator is not trivial in non-Lorentz invariant backgrounds 
may  in principle provide a powerful tool to derive identities 
involving matrix elements of operators and the spectrum in weakly as well as strongly coupled QFTs.
Consider for example a QFT on the cylinder ${\mathbb R}^{D-1}\times S^1$; the K\"allen-Lehmann-like 
representation for the commutator vev of two local operators reads
\beq \label{eq:sr1}
\langle 0 | [{\cal O}^\dagger (x,y), {\cal O} (0,0)] |0\rangle=
\sum_{n}\sum_{\lambda} \left |\langle \lambda;m_\lambda(n),n| {\cal O}(0,0) |0\rangle\right|^2 
e^{i n y/R} \Delta^{D-1}(x;m_\lambda(n)) \,,
\eeq
where the formal sum in $\lambda$ runs over all the states of the theory with
 KK momentum $n/R$ (and includes the integral on the corresponding phase-space), 
$m_\lambda(n)$ is the invariant $(D-1)$-dimensional mass
and $\Delta^{D-1}(x;m_\lambda(n))$ is the $(D-1)$-dimensional commutator vev of a free field with mass  $m_\lambda(n)$.
In principle we can now use the fact that we know that the commutator vanishes outside the light cone to get
a non trivial relation. For instance we can integrate eq.~(\ref{eq:sr1}) with a generic function $f(t,y)$ outside the light cone (i.e.~on the diamond 
$\diamondsuit\equiv\{t,y:t^2<(y+n R)^2, \forall n\in {\mathbb Z}\}$) 
and get
\beq
\sum_{\lambda}\sum_{n} \left |\langle \lambda;m_\lambda(n),n| {\cal O}(0,0) |0\rangle\right|^2 
\frac{1}{m_\lambda(n)} \int_\diamondsuit\, dy\, dt\, \cos(2\pi n y/R)\, \sin(m_\lambda(n)t)\, f(t,y)=0 \,;
\eeq
in order not to get contributions from delta-like singularities localized on the light-cone we have to restrict to functions $f$ that vanish on the light cone.
The above relation gives sum rules involving the matrix elements and the spectrum of the theory and 
that are non-trivial especially when the theory is strongly interacting.
It would be interesting to study whether such relations could be used 
in similar contexts like extra-dimensional models or QCD on the lattice.
For instance it was noticed in \cite{kachru} that in string compactifications the mass of the
lightest modulus is always parametrically smaller than the KK scale. One might 
expect that such inequality follows from microcausality in the compactified space.

\section*{Acknowledgments}
It is a pleasure to thank Nima Arkani-Hamed for his collaboration throughout this work, and Riccardo Rattazzi for useful discussions and for pointing out 
that microcausality should follow from the Heisenberg-picture 
equations of motion.
We also thank Tim Hollowood and Graham Shore for correspondence.
The work of Enrico Trincherini is supported by an INFN postdoctoral fellowship.

%%%%%%%%%%%%%%%%%%%%%%%%%%%%%%%%%
%%%%%%%%%%%%%%%%%%%%%%%%%%%%%%%%%

\end{document}